\documentclass[conference]{IEEEtran}
\IEEEoverridecommandlockouts
\usepackage{url}
\usepackage{xurl}
\usepackage{cite}
\usepackage{amsmath,amssymb,amsfonts}
\usepackage{graphicx}
\usepackage{textcomp}
\usepackage{xcolor}
\usepackage{hyperref}
\usepackage{stfloats}
\usepackage{booktabs}
\DeclareMathOperator{\clip}{clip}
\DeclareMathOperator{\hnrop}{HNR}
\newcommand{\un}[1]{\,\mathrm{#1}}
\newcommand{\Nm}{\ensuremath{\mathrm{N{\cdot}m}}}

\newcommand{\fs}{f_{\mathrm{s}}}
\newcommand{\preemph}{\alpha}
\newcommand{\frameidx}{t}
\newcommand{\nframes}{T}
\newcommand{\freq}{f}
\newcommand{\win}{\ell}
\newcommand{\spec}{X}
\newcommand{\specw}{\tilde{X}}
\newcommand{\gauss}{G}
\newcommand{\sigw}{\sigma_{\mathrm{w}}}
\newcommand{\sigc}{\sigma_{\mathrm{c}}}

\newcommand{\speed}{\omega}
\newcommand{\speedhat}{\hat{\omega}}
\newcommand{\accel}{\dot{\omega}}
\newcommand{\accelhat}{\dot{\hat{\omega}}}
\newcommand{\frot}{f_{\mathrm{rot}}}
\newcommand{\frothat}{\hat{f}_{\mathrm{rot}}}
\newcommand{\order}{k}
\newcommand{\orderset}{\mathcal{K}}
\newcommand{\ncyl}{n}
\newcommand{\cand}{r}
\newcommand{\candset}{\mathcal{R}}
\newcommand{\binw}{\delta}

\newcommand{\cell}{\mathrm{cell}}
\newcommand{\combkern}{\psi}
\newcommand{\comb}{E}
\newcommand{\post}{P}

\newcommand{\vel}{v}
\newcommand{\stiffa}{\kappa_{\mathrm{a}}}

\newcommand{\torque}{\tau}
\newcommand{\torquehat}{\hat{\tau}}
\newcommand{\torquewot}{\tau_{\mathrm{WOT}}}
\newcommand{\torquedrag}{\tau_{\mathrm{drag}}}
\newcommand{\torqueload}{\tau_{\mathrm{L}}}
\newcommand{\inertia}{\Theta}
\newcommand{\load}{\lambda}
\newcommand{\loadhat}{\hat{\lambda}}
\newcommand{\loadstate}{u}
\newcommand{\loadstatehat}{\hat{u}}
\newcommand{\feat}{\mathbf{z}}
\newcommand{\bandprof}{B}
\newcommand{\hnr}{\hnrop}
\newcommand{\lambins}{D}
\newcommand{\lambinc}{\lambda}
\newcommand{\siglam}{\sigma_{\lambda}}

\newcommand{\papertitle}{RPM and Torque Estimation from Audio: Recovering Engine Operating Parameters from Harmonic Structure and Rotational Dynamics}
\newcommand{\paperkeywords}{engine speed estimation, torque estimation, order analysis, harmonic comb, Viterbi decoding, active sound design}
\newcommand{\paperabstract}{Engine Order Enhancement in Active Sound Design and data-driven engine synthesis rely on speed and torque, which recordings without telemetry lack. We recover both: speed with a front end that computes harmonic order energy on a logarithmic speed grid and a decoder, motivated by the drivetrain equation, that charges changes of the relative rate of speed change. As absolute torque is not identifiable from audio alone, we estimate a load intensity between the engine's drag floor and full-load ceiling from order levels and harmonic-to-noise ratio, rescalable to the expected torque range. On three engines, loudness, the usual load proxy, explains 0.18 to 0.52 of the torque variance (R\textsuperscript{2}), an inertial model 0.23 to 0.68, and order band levels with speed 0.86 to 0.95. Our front end and decoder find the speed more often than four pitch trackers given the harmonic order; the load intensity (0 to 1) has a mean absolute error of 0.09 to 0.12, with the torque sign right on 90 to 95\,\% of frames.}

\newcommand{\captureCellBins}{6.5}
\newcommand{\captureHalfTileKtwelve}{0.998}
\newcommand{\captureHalfPointKtwelve}{0.354}

\begin{document}

\title{\papertitle}

\author{\IEEEauthorblockN{Robin Doerfler}
\IEEEauthorblockA{\textit{Audio Research \& Development} \\
\textit{Impulse Audio Lab GmbH}\\
Munich, Germany \\
robin.doerfler@impulse-audio-lab.com}
\and
\IEEEauthorblockN{Beat Rossmy}
\IEEEauthorblockA{\textit{Audio Research \& Development} \\
\textit{Impulse Audio Lab GmbH}\\
Munich, Germany \\
beat.rossmy@impulse-audio-lab.com}
}

\maketitle

\begin{abstract}
\paperabstract
\end{abstract}


\section{Introduction}
\label{sec:introduction}

Engine speed and load set the operating state of a combustion engine and shape its sound. Active Sound Design (ASD) maps the amplitudes of engine orders over speed and torque~\cite{doerflerGradientBasedLearning2026}, data-driven engine synthesis learns the sound of a recorded engine as a function of such control signals~\cite{lobatoMotor2Synth2025, choiDDSPNeuralVehicle2026, doerflerPhysicsInformedNeural2026}, and order analysis needs the speed as the reference of every order it extracts~\cite{peetersReviewComparisonTacholess2019}. Both signals come from a tachometer, a dynamometer or the Controller Area Network (CAN) bus of the vehicle~\cite{choiDDSPNeuralVehicle2026}, while recordings in archives, sample libraries and online video carry neither~\cite{lobatoMotor2Synth2025}. Synthesis models trained on such recordings can fall back on the loudness of the signal as their second control~\cite{engelDDSPDifferentiableDigital2020, lobatoMotor2Synth2025}, but as a stand-in for the load it is ambiguous: a run-up and an overrun can reach the same level, and one level means a different effort on a different engine.

The speed is present in the positions of the signal's harmonic components, the engine orders, yet an analysis has to separate orders a few hertz apart at idle while following them through run-ups in which they sweep tens of hertz within a fraction of a second (Section~\ref{subsec:orders}). General pitch estimators return one periodicity per frame~\cite{mauchPYIN2014, kimCREPE2018, riouPESTO2023}, which gives the speed only once the order of that periodicity is known, and that order depends on the cylinder count. Tacholess order tracking takes it from the known construction of the machine, such as its gear ratios~\cite{peetersReviewComparisonTacholess2019}. The absolute torque in \Nm{} of an unknown engine is not identifiable from the sound (Section~\ref{subsec:load_torque}). We therefore estimate how hard the engine works against its own capacity, as a \emph{load intensity} $\load$ between the engine's drag floor and its full-load ceiling, and convert it to torque where these limits are known.

\begin{figure*}[t]
\centering
\includegraphics[width=\textwidth]{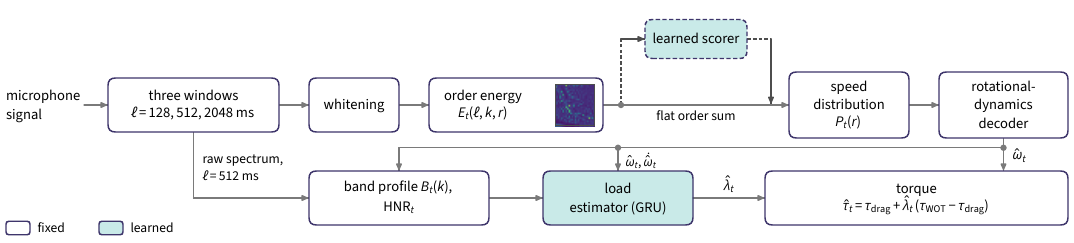}
\input{figures/fig_chain.caption}
\label{fig:chain}
\end{figure*}

Fig.~\ref{fig:chain} shows the estimator. A front end extracts the \emph{order energy} from each spectral frame: the spectral energy at the positions of the engine orders, computed for every candidate speed on a logarithmic grid from 600 to $9000\un{rpm}$. A decoder that imposes the rotational dynamics of the engine turns the frames into one speed trajectory, and a load estimator reads the order levels of the raw spectrum at the decoded speed. The complete system~\cite{doerflerSonoDyno2026} adds a learned scorer between the front end and the decoder, which reads the speed together with the cylinder count; here the front end and the decoder are also evaluated on their own, without any learned stage.

The contributions are:
\begin{itemize}
  \item a domain-specific front end of harmonic order energy on a logarithmic speed grid, with kernels that tile the frequency axis and three window lengths whose roles at idle and in run-ups we measure (Section~\ref{sec:order_energy});
  \item a decoder motivated by the drivetrain equation, which charges changes of the relative rate of speed change and leaves steady ramps free (Section~\ref{sec:decoder});
  \item the engine-normalised load intensity, a comparison of the load cues in the sound, and the estimation of the load intensity from the order band profile at the decoded speed (Sections~\ref{sec:signal} and~\ref{sec:load}).
\end{itemize}

\section{Engine Sound, Speed and Load}
\label{sec:signal}

\subsection{Orders and the Resolution Conflict}
\label{subsec:orders}

Each of the $\ncyl$ cylinders of a four-stroke engine releases one exhaust pressure pulse per working cycle of two crankshaft revolutions~\cite{jonesModellingExhaustNoise1984}, so the sound has components at the orders $\order \in \{\tfrac12, 1, \tfrac32, \dots\}$, at $\order\frot$ with $\frot = \speed/60$ and $\speed$ the engine speed in rpm~\cite{wangVehicleNoiseVibration2010}. Equal cylinders firing at equal intervals radiate at the firing order $\ncyl/2$ and its multiples alone; differences between the cylinders fill the orders in between~\cite{wangVehicleNoiseVibration2010}. The firing frequency $\ncyl\speed/120$ depends on count and speed only through their product: twice the cylinders at half the speed give the same frequency. Summed over the orders, the order energy of a candidate speed measures how much of the spectrum the harmonics of that speed would explain, and this sum peaks at the true speed and again an octave from it. The complete system resolves these octave peaks by reading the count together with the speed~\cite{doerflerSonoDyno2026}.

At $600\un{rpm}$, the lowest speed considered here, neighbouring half orders lie $5\un{Hz}$ apart, and the main lobe of a Blackman window, $\pm 3/\win$ to its first zeros, separates them only from a length $\win$ of about $600\un{ms}$. In a run-up at $1500\un{rpm/s}$, order $\order$ moves by $25\,\order\un{Hz}$ per second, so that within $600\un{ms}$ order 4 sweeps $60\un{Hz}$ and order 12 sweeps $180\un{Hz}$.

\subsection{Load and Torque}
\label{subsec:load_torque}

Even at constant engine speed, the load changes the pulses~\cite{doerflerPhysicsInformedNeural2026} and with them the energy distribution across the orders~\cite{doerflerGradientBasedLearning2026}; full load raises the upper part of the spectrum most~\cite{tekinExperimentalInvestigationEffects2022}. With the fuel cut off in overrun, nothing burns and the engine only pumps air~\cite{doerflerPhysicsInformedNeural2026}. From 1500 to $3000\un{rpm}$ in our real recordings, high load ($\load > 0.6$, Eq.~\eqref{eq:lambda}) raises the count of orders standing at least $6\un{dB}$ above the floor between them by a factor of 1.3 to 1.8 over overrun on six of the seven engines measured.

Reduced to the crankshaft, the drivetrain follows
\begin{equation}
  \inertia\,\tfrac{2\pi}{60}\,\accel = \torque - \torqueload,
  \label{eq:drivetrain}
\end{equation}
with $\torque$ the engine torque, $\torqueload$ the load torque, $\inertia$ the effective inertia of engine and vehicle seen at the crankshaft through the engaged gear, and $2\pi/60$ the conversion of the acceleration $\accel$ from rpm per second to $\un{rad/s^2}$. From the harmonic structure of the sound we estimate the speed $\speed$ and, from its course, its rate of change $\accel$ (Sections~\ref{sec:order_energy} and~\ref{sec:decoder}). The inertia $\inertia$ and the load torque $\torqueload$, however, differ between vehicles with their mass and change with the gear and the slope of the road, so the audio does not reveal them. The rate of change thus only shows whether the engine torque exceeds the load torque and leaves its size open: at a steady speed the engine torque matches the load torque, small on a flat road and large uphill. Since the net torque varies slowly over a frame, so does the rate of speed change, which the decoder of Section~\ref{sec:decoder} builds on; the load estimator takes the rate of change as one input (Section~\ref{sec:load}).

The torque also has to be read against the capacity of the engine: two engines of our corpus peak at 304 and $803\un{\Nm}$, so $300\un{\Nm}$ saturates the first and uses less than 40\,\% of the second. We therefore describe the load as the relative effort of the engine between its drag floor $\torquedrag(\speed)$, the torque with the fuel cut off in overrun, and its full-load ceiling $\torquewot(\speed)$ at wide-open throttle, and define this position as the load intensity
\begin{equation}
  \load = \clip\!\left(\frac{\torque - \torquedrag(\speed)}{\torquewot(\speed) - \torquedrag(\speed)},\,0,\,1\right),
  \label{eq:lambda}
\end{equation}
with both curves fitted to the labelled torque of each engine, so that $\load = 0$ is the deepest overrun and $\load = 1$ full load at any speed of any engine, and one estimator can learn $\load$ across engines of any size.

\subsection{Load Cues in the Sound}
\label{subsec:cues}

We compared three groups of features by how much of the torque they explain: (1)~the level and the A-weighted loudness, (2)~an inertial model built on the drivetrain equation, and (3)~the energies in bands around the orders. Each group, computed for this comparison only, fed a model of the torque in \Nm{} that was fitted on recordings of three engines and scored on recordings left out of the fit, with the speed taken from the labels as if estimated perfectly. (1)~A linear regression on the level and the loudness, the most direct cues, explains 0.18 to 0.52 of the torque variance ($\mathrm{R}^2$); on two of the three engines, the speed and its rate of change alone explain more (0.46 to 0.77). It tells overrun, cruise and acceleration apart on at most half of the frames. (2)~The inertial model, with one fitted inertia and a load torque quadratic in $\speed$, explains 0.23 to 0.68, and at most 0.36 on the steadiest quarter of the frames, where the rate of speed change says little about the load. (3)~A small multilayer perceptron on the energies in bands of $\pm 3\,\%$ around the orders, with the level, the spectral tilt and the speed, explains 0.86 to 0.95 and tells the three states apart on 80 to 88\,\% of the frames. Section~\ref{sec:load} therefore reads the load from the order levels at the estimated speed.

\section{Harmonic Order Energy}
\label{sec:order_energy}

\subsection{Spectra and Whitening}
\label{subsec:whitening}

The audio is resampled to $\fs = 8\un{kHz}$, which holds order 12 up to $9000\un{rpm}$, and pre-emphasised with $\preemph = 0.97$. Every $64\un{ms}$, a frame $\frameidx$ is cut out by three centred Blackman windows of length $\win$~=~128, 512 and $2048\un{ms}$. Zero-padding the two shorter windows fourfold gives bins of $0.488\un{Hz}$ at 512 and $2048\un{ms}$ and of $1.95\un{Hz}$ at $128\un{ms}$. We write $\spec_\frameidx(\freq, \win)$ for each magnitude spectrum.

Because the spectral envelope changes with the microphone position while the order frequencies do not, we whiten each frame to $\specw_\frameidx(\freq, \win)$ by subtracting its smoothed log spectrum, which keeps the order lines and makes them comparable across microphone positions,
\begin{equation}
  \log \specw_\frameidx(\freq, \win) = \log \spec_\frameidx(\freq, \win) - \sum_{\freq'} \gauss(\freq - \freq', \sigw)\, \log \spec_\frameidx(\freq', \win),
  \label{eq:whitening}
\end{equation}
with $\gauss$ a unit-mass Gaussian of width $\sigw = 29\un{Hz}$, followed by a normalisation of $\specw_\frameidx$ to its maximum.

\subsection{A Logarithmic Speed Grid}
\label{subsec:grid}

The candidate speeds $\cand$ form a grid $\candset$ of 512 points with logarithmic spacing between 600 and $9000\un{rpm}$: neighbours differ by $\binw = 0.00765$ octave, 0.531\,\% of the speed, so every octave holds 131 candidates. The step thus grows with the speed as the spacing of the orders does, from $3.2\un{rpm}$ at $600\un{rpm}$ to $48\un{rpm}$ at $9000\un{rpm}$. On this grid a relative change of speed moves all orders by one common number of candidates, so a ramp at a given relative rate, or an octave error, has one shape at every speed and every order.

\subsection{Tiling Kernels}
\label{subsec:tiling}

For order $\order$ and candidate $\cand$, the kernel $\combkern(\freq, \win, \order, \cand)$ averages the Gaussian line $\gauss$ at the frequency $\order\cand'/60$ of order $\order$ over every speed $\cand'$ in the cell of $\cand$. This cell, $\cell(\cand)$, holds the speeds closer to $\cand$ than to its neighbouring candidates. Scaled to unit mass over the bins of window $\win$, the kernel is
\begin{equation}
  \combkern(\freq, \win, \order, \cand) \propto \int_{\cell(\cand)} \gauss\!\left(\freq - \frac{\order \cand'}{60}, \sigc\right) d\cand',
  \label{eq:kernel}
\end{equation}
with the Gaussian width $\sigc$ set to 1.5 bins of that window. The kernel is thus a box over the band that order $\order$ occupies for the speeds of the cell, with edges blurred by the Gaussian. Weighting the whitened spectrum $\specw_\frameidx(\freq, \win)$ with the kernel gives the order energy,
\begin{equation}
  \comb_\frameidx(\win, \order, \cand) = \sum_{\freq} \combkern(\freq, \win, \order, \cand)\, \specw_\frameidx(\freq, \win),
  \label{eq:orderenergy}
\end{equation}
over the 24 orders $\orderset = \{\tfrac12, 1, \tfrac32, \dots, 12\}$, so that each frame holds $3 \times 24 \times 512$ values.

\begin{figure*}[t]
\centering
\includegraphics[width=\textwidth]{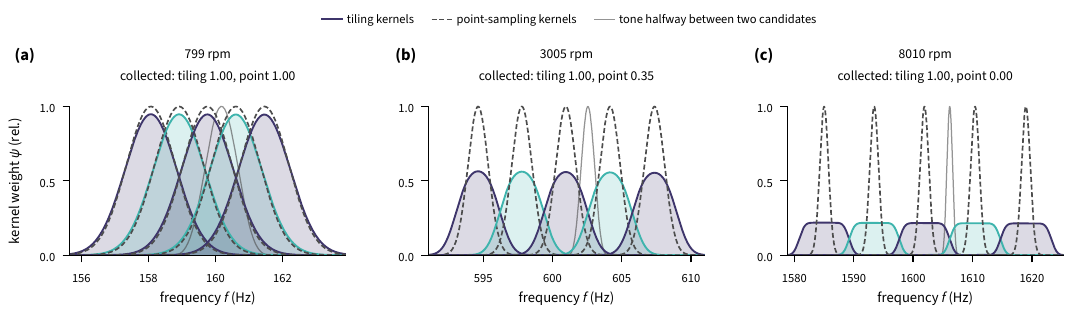}
\input{figures/fig_capture.caption}
\label{fig:capture}
\end{figure*}

Because the cells of neighbouring candidates abut, so do their bands at every order: the kernels tile the frequency axis (Fig.~\ref{fig:capture}), such that a harmonic at any speed within a cell lies under that cell's kernel. The width of a band, $\order\,\lvert\cell(\cand)\rvert/60 \approx 0.0053\,\order\cand/60$ in hertz, grows with the order and with the speed, in proportion to the distance a harmonic moves for a given relative change of speed: from far below one bin for order 1 at $600\un{rpm}$ to twenty bins of the long window for order 12 at $9000\un{rpm}$. A point-sampling kernel, a Gaussian of fixed width at $\order\cand/60$, collects as much wherever the main lobe of the window is wider than the band. In the 128 and $512\un{ms}$ windows the main lobe, $\pm 23$ and $\pm 5.9\un{Hz}$ to its first zeros, is wider than every band, and the two kernels collect alike. In the $2048\un{ms}$ window, which resolves the orders at idle, one step of the grid moves order 12 at $3000\un{rpm}$ by \captureCellBins{} bins, more than the main lobe covers: of a tone halfway between two candidates, the point-sampling kernels collect only \captureHalfPointKtwelve{} of the energy that a tone on a candidate yields, the tiling kernels \captureHalfTileKtwelve{} (Fig.~\ref{fig:capture}b). On recordings, both kernels find the speed on the same share of frames, within two percentage points, so tiling closes the gap between candidates at no cost in accuracy.

A ramp that changes the speed by a given factor within the window sweeps order $\order$ over a band $\order$ times as wide as order 1. The kernels widen in the same proportion, so the smeared harmonic covers the same number of candidates at every order, and the orders of a ramp stay aligned when they are summed. Only the main lobe of the window, fixed in hertz, blurs the low orders over more candidates than the high ones.

\subsection{Three Windows}
\label{subsec:windows}

\begin{figure*}[t]
\centering
\includegraphics[width=\textwidth]{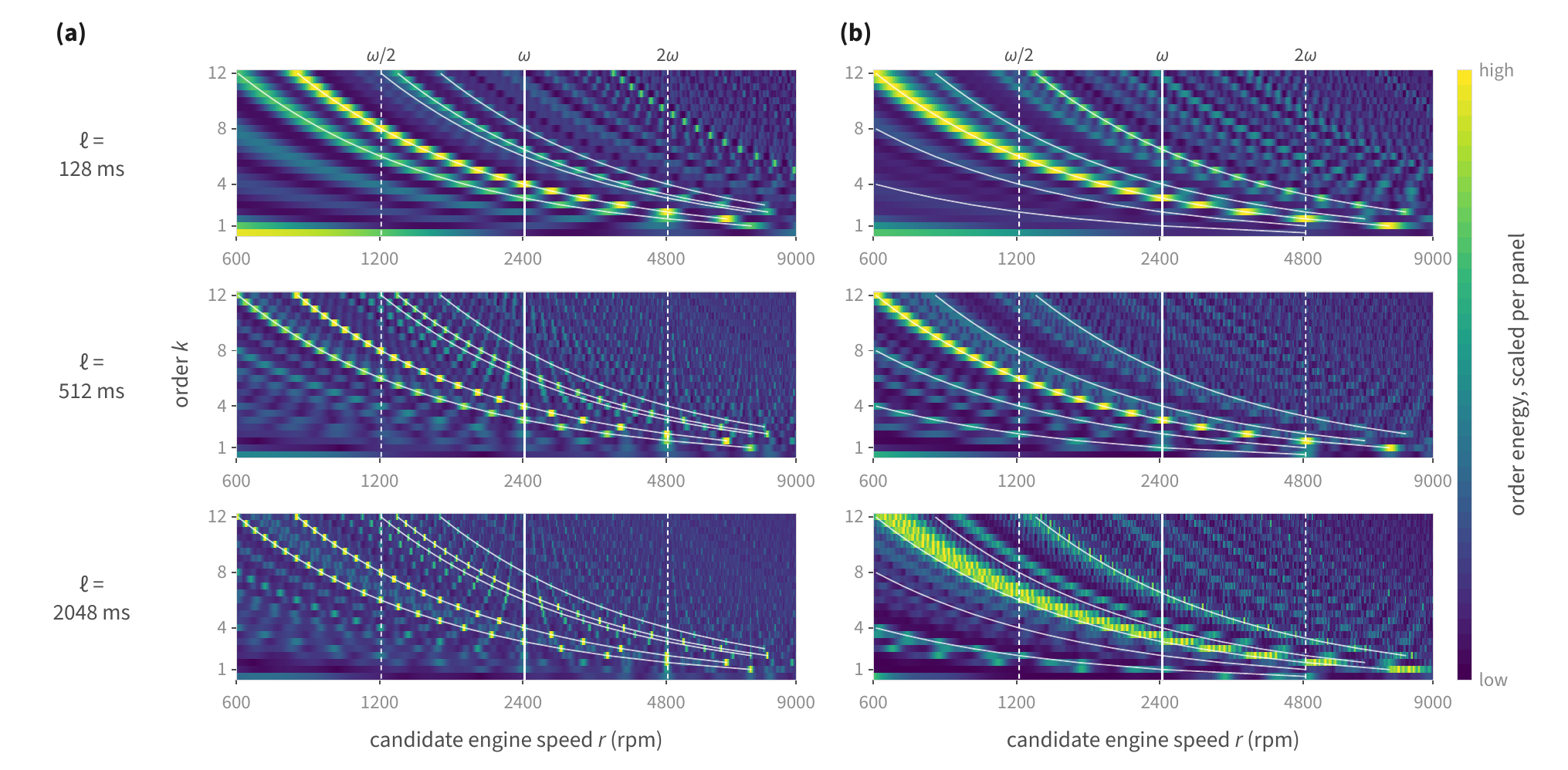}
\input{figures/fig_views.caption}
\label{fig:views}
\end{figure*}

Picking in each frame the candidate $\cand$ whose order energy, summed over the 24 orders with equal weight, is largest (the \emph{flat order sum}) finds the speed within 5\,\% with the $2048\un{ms}$ window on 84\,\% of the nearly steady frames, but on only 17\,\% in ordinary run-ups, where the $512\un{ms}$ window keeps 66\,\% (Table~\ref{tab:windows}). At standstill, the $512\un{ms}$ window reads twice the speed on 40\,\% of the steady frames and the $2048\un{ms}$ window on 11\,\%. The orders of the double speed fall on the integer orders of the true speed, so only the half orders tell the two apart. At this engine's idle they lie $7\un{Hz}$ apart, barely separated by the $512\un{ms}$ main lobe (about $6\un{Hz}$ half-width) and fully by the $2048\un{ms}$ one ($1.5\un{Hz}$).
\begin{table}[t]
\caption{Flat order sum of each window: share of frames within 5\,\% of the true speed, by rate of speed change, on the held-out sessions (Section~\ref{sec:data}).}
\label{tab:windows}
\centering
\footnotesize
\begin{tabular}{@{}lrrrr@{}}
\toprule
Speed change (rpm/s) & Frames & $128\un{ms}$ & $512\un{ms}$ & $2048\un{ms}$ \\
\midrule
0 to 100        & 37\,927 & 0.13 & 0.67 & \textbf{0.84} \\
100 to 500      & 55\,176 & 0.36 & \textbf{0.88} & 0.68 \\
500 to 1500     & 20\,630 & 0.45 & \textbf{0.66} & 0.17 \\
above 1500      & 10\,855 & 0.15 & \textbf{0.20} & 0.11 \\
\midrule
all             & 124\,588 & 0.29 & \textbf{0.72} & 0.59 \\
\bottomrule
\end{tabular}
\end{table}

The $128\un{ms}$ window falls behind the $512\un{ms}$ window, since its main lobe merges neighbouring orders, yet its ridge keeps its width through a run-up, where the $2048\un{ms}$ ridge smears (Fig.~\ref{fig:views}b).

\section{Rotational-Dynamics Decoder}
\label{sec:decoder}

The decoder reads a speed distribution $\post_\frameidx(\cand)$ over the candidates in every frame, which sums to one and states how strongly the frame supports each candidate: the mean order energy over the three windows, summed over the orders and normalised, which has no learned parameter and keeps the octave peaks of Section~\ref{subsec:orders}, or the distribution of the learned scorer of the complete system, which also reads the cylinder count~\cite{doerflerSonoDyno2026}.

The acceleration of the crankshaft follows the net torque (Eq.~\eqref{eq:drivetrain}), and the net torque changes slowly against a $64\un{ms}$ frame, apart from the moment of a gear change. A speed path that keeps its rate of change from frame to frame is therefore physical, while a path that jitters between neighbouring candidates or jumps to an octave peak is not.

\begin{figure}[t]
\centering
\includegraphics[width=\columnwidth]{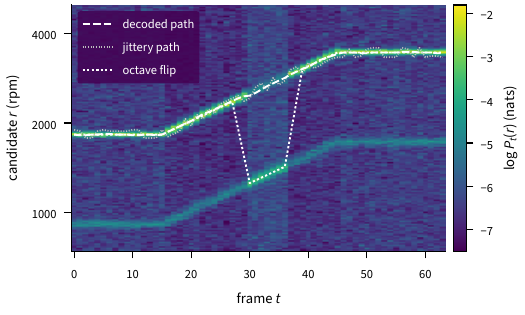}
\input{figures/fig_decoder.caption}
\label{fig:decoder}
\end{figure}

On the logarithmic grid, the slope of a speed path is the path velocity $\vel_\frameidx = (\log_2 \cand_\frameidx - \log_2 \cand_{\frameidx-1})/\binw$, the number of candidates the speed moves per frame, which measures its relative rate of change. For a path $(\cand_1, \dots, \cand_\nframes)$ over the $\nframes$ frames of a recording, the decoder finds
\begin{equation}
  \arg\max_{\cand_1, \dots, \cand_\nframes} \sum_{\frameidx=1}^{\nframes} \log \post_\frameidx(\cand_\frameidx) - \stiffa \sum_{\frameidx=3}^{\nframes} \big((\vel_\frameidx - \vel_{\frameidx-1})\,\binw\big)^2
  \label{eq:decoder}
\end{equation}
by a Viterbi search~\cite{viterbi1967} over the states (candidate, path velocity), with $\stiffa = 512$ the weight on the change of the path velocity. A straight ramp, whose slope stays constant, costs nothing at any rate; each change of slope costs in proportion to its square. A first-order decoder, which charges the slope itself, lags behind a sustained run-up.

Since $(\vel_\frameidx - \vel_{\frameidx-1})\,\binw$ equals the second difference of $\log_2 \cand_\frameidx$, measured in octaves per frame squared, the same relative acceleration produces the same change of slope at every speed: a run-up that gains 10\,\% per second costs as much at $800\un{rpm}$ as at $6000\un{rpm}$, although it gains far more rpm at the higher speed. One weight $\stiffa$ therefore holds over the whole speed range. The slope is bounded at $\lvert \vel_\frameidx \rvert \le 48$ candidates per frame, a change of the speed by up to 29\,\% within one $64\un{ms}$ frame, so that even the $1500\un{rpm/s}$ run-up of Section~\ref{subsec:orders} stays inside the bound at $600\un{rpm}$, where it changes the speed by 16\,\% per frame (Fig.~\ref{fig:decoder}). The decoded path moves in whole steps of the grid, 0.531\,\% of the speed; the mean of the candidates within $\pm 3$ steps of it, each weighted by $\post_\frameidx(\cand)$, interpolates between the steps and gives the speed estimate $\speedhat_\frameidx$.

\section{Load-Intensity Estimation}
\label{sec:load}

\subsection{Order Levels at the Decoded Speed}
\label{subsec:features}

The load acts on the levels of the orders and on the floor between them (Section~\ref{subsec:load_torque}); the normalisation and the whitening of Section~\ref{subsec:whitening} remove the absolute level and the envelope. The load estimator therefore reads the power spectrum $\lvert \spec_\frameidx(\freq, 512) \rvert^2$ of the $512\un{ms}$ window before both, at the decoded rotation frequency $\frothat = \speedhat_\frameidx/60$, in 27 values per frame:
\begin{itemize}
  \item \textbf{Band profile:} $\bandprof_\frameidx(\order)$, the power in dB within $\pm 3\,\%$ of $\order\,\frothat$, for each of the 24 orders of $\orderset$;
  \item \textbf{Harmonic-to-noise ratio:} $\hnr_\frameidx$, the ratio in dB of the band energy of orders up to 8 to the energy between those bands;
  \item \textbf{Speed and its rate of change:} $\speedhat_\frameidx$ and $\accelhat_\frameidx$, the central difference of the decoded speed averaged over five frames.
\end{itemize}
The bands of $\pm 3\,\%$ hold the harmonic of every order while the decoded speed is within 3\,\% of the true one. Standardised, the 27 values form the input $\feat_\frameidx$ of the estimator.

\subsection{Recurrent Estimator and Readout}
\label{subsec:head}

The estimator is a two-layer gated recurrent unit (GRU)~\cite{choGRU2014} with 128 units per layer. It reads $\feat_1, \dots, \feat_\frameidx$ and outputs per frame a distribution $\lambins_\frameidx(j)$ over 40 equally wide bins $j$ of $\load$, centred at $\lambinc(j)$, and the probability $\loadstatehat_\frameidx$ of positive torque.
For one and the same order profile, the distribution often holds two modes, at overrun and under load. Its centre of mass,
\begin{equation}
  \loadhat_\frameidx = \sum_{j=1}^{40} \lambins_\frameidx(j)\, \lambinc(j),
  \label{eq:loadhat}
\end{equation}
is the estimate. The peak would jump between the modes from frame to frame: it explains 0.63 of the variance of $\load$ on the held-out sessions (Section~\ref{sec:data}), the centre of mass 0.76.

\subsection{Training and Conversion to Torque}
\label{subsec:training}

The training loss is the cross-entropy between $\lambins_\frameidx$ and a Gaussian of width $\siglam = 1$ bin around the labelled $\load_\frameidx$, with 0.3 times a binary cross-entropy added for $\loadstatehat_\frameidx$ against the load state $\loadstate_\frameidx$ (1 for positive torque, 0 in overrun). The estimator is trained on features at the decoded speed of the complete system, the speed it receives in use. Since the absolute level depends on the microphone and the recording chain, training offsets all 24 band levels of a session by one common random gain within $\pm 12\un{dB}$.

Where the drag floor and the full-load ceiling of the engine are known, the estimate converts to torque by inverting Eq.~\eqref{eq:lambda},
\begin{equation}
  \torquehat_\frameidx = \torquedrag(\speedhat_\frameidx) + \loadhat_\frameidx \left(\torquewot(\speedhat_\frameidx) - \torquedrag(\speedhat_\frameidx)\right).
  \label{eq:torque}
\end{equation}
For a session with several microphones, each frame takes the mean of $\loadhat_\frameidx$ and of $\loadstatehat_\frameidx$ over the microphones.

\section{Data and Evaluation}
\label{sec:data}

The corpus holds 116.7\,h of real and simulated engine audio at 48\,kHz, labelled sample by sample with engine speed and torque from the vehicle telemetry or the simulation. The real recordings, 41.9\,h summed over microphones, cover 8 engines with four, six and eight cylinders in 538 sessions, mostly on chassis dynamometers, with 2 to 14 microphones per session out of 14 positions in the engine bay, near the exhaust and inside the cabin. Engine simulations and procedural resyntheses of two of the real engines contribute the other 74.8\,h~\cite{doerflerProceduralEngineSoundsDataset2025, doerflerAnalysisDrivenProceduralGeneration2026}.

The split into training, validation and test material is grouped by recording session. The real validation recordings are the \emph{held-out sessions} of a real eight-cylinder engine trained on its other sessions and the \emph{unseen-type engine}, a real four-cylinder whose type training never saw, recorded with another setup; simulated engines absent from training complete the validation. The \emph{test engine} is a real eight-cylinder engine whose type training knows only from one other engine of the same type.

We call one microphone's signal over one session a recording. For the speed we report c5 and c2, the shares of labelled frames (600 to $9000\un{rpm}$) whose estimate lies within 5 and 2\,\% of the true speed, and the mean absolute relative error (RMAE). A recording is \emph{broken} if its c5 is below 0.2. The labels are sampled every $64\un{ms}$ and step within one frame at gear shifts and free revs, which any reading through an analysis window rounds through, a correct one included: the labels themselves, smoothed over $192\un{ms}$, reach only c2 0.98 and c5 0.99 on the real validation recordings. For the load intensity we report the mean absolute error (MAE) of $\loadhat_\frameidx$ and the explained variance $\mathrm{R}^2$ of $\load_\frameidx$, pooled over running frames, and the state accuracy, the share of frames where $\loadstatehat_\frameidx \ge 0.5$ matches the torque sign.

\section{Results}
\label{sec:results}

\subsection{Engine Speed}
\label{subsec:results_speed}

Table~\ref{tab:speed} compares the estimates on the real validation recordings. The general pitch and periodicity trackers YIN~\cite{decheveigneYIN2002}, pYIN~\cite{mauchPYIN2014}, CREPE~\cite{kimCREPE2018} and PESTO~\cite{riouPESTO2023} track one periodicity, which we convert to a speed with the harmonic order that suits each engine best, chosen on the labels. The flat order sum is the speed distribution of Section~\ref{sec:decoder} without a learned stage, decoded with Eq.~\eqref{eq:decoder}. Weighting the firing orders of the true cylinder count gives the sum the engine knowledge the trackers receive. The complete system~\cite{doerflerSonoDyno2026} reads the count from the audio.

\begin{table}[t]
\caption{Speed estimates on 163 of the 334 real validation recordings (67\,827 labelled frames).}
\label{tab:speed}
\centering
\footnotesize
\begin{tabular}{@{}lrrrr@{}}
\toprule
System & c5 & c2 & Broken & RMAE (\%) \\
\midrule
YIN, best order                & 0.279 & 0.275 & 0.552 & 81.1 \\
pYIN, best order               & 0.258 & 0.253 & 0.626 & 78.8 \\
CREPE, best order              & 0.378 & 0.342 & 0.337 & 206.4 \\
PESTO, best order              & 0.616 & 0.569 & 0.172 & 63.8 \\
\midrule
Flat order sum, no learning     & 0.718 & 0.709 & 0.301 & 19.9 \\
Firing orders of the true count & 0.775 & 0.764 & 0.203 & 20.0 \\
Complete system                & \textbf{0.957} & \textbf{0.948} & \textbf{0.006} & \textbf{2.4} \\
\bottomrule
\end{tabular}
\end{table}

The order energy and the decoder alone, with no learned parameter and no knowledge of the engine, place more frames within 5\,\% of the true speed than any tracker handed the harmonic order (0.72 against 0.62), and within 2\,\% as well (0.71 against 0.57). Below $3000\un{rpm}$ many of their errors are octave errors, since a flat order sum does not know which octave peak is the engine; on all real validation recordings they find the speed on 45\,\% of the idle frames and on 60\,\% of ordinary run-up frames. From $\stiffa = 128$ to $2048$, and with a first-order decoder, their c5 on the real validation recordings stays between 0.71 and 0.73: the flat order sum itself limits the estimate.

The complete system, which reads the cylinder count and with it the octave, reads the speed within 2\,\% on 95\,\% of the frames, inside the $\pm 3\,\%$ that the load bands tolerate (Section~\ref{subsec:features}); Section~\ref{subsec:results_load} evaluates the load at its speed.

On the speed distributions of two other configurations of the learned scorer, we varied only the decoder. On all validation recordings, raising the weight of a first-order decoder from 64 to 1024 lowers the share of frames above $1500\un{rpm/s}$ within 5\,\% from 0.24 to 0.11: a stiff path lags behind every run-up. On the real validation recordings, charging the change of the path velocity in its place raises c5 from 0.900 to 0.904 and those frames from 0.15 to 0.21, and lowers the broken share from 0.054 to 0.046. With the second, raising the velocity bound from 12 to 48 candidates per frame lifts those frames from 0.54 to 0.63.

\subsection{Load Intensity}
\label{subsec:results_load}

With the speed estimated by the complete system, the load intensity explains less of the variance per microphone on the test engine than on the held-out sessions (0.64 against 0.76), and the session average over the microphones recovers most of the gap (0.78 against 0.82; Table~\ref{tab:load}). The torque sign is correct on 90 to 95\,\% of the frames of both, so the estimate tells traction from overrun, the two states a level alone confuses (Section~\ref{sec:introduction}).

\begin{table}[t]
\caption{Load intensity with the speed estimated by the complete system, on the validation and test material, per microphone and as session average.}
\label{tab:load}
\centering
\footnotesize
\begin{tabular}{@{}llrrr@{}}
\toprule
Material & Read & $\mathrm{R}^2$ & MAE & State \\
\midrule
\multicolumn{5}{@{}l}{\emph{Validation}} \\
Held-out sessions         & per microphone & 0.756 & 0.090 & 0.905 \\
                          & session average & 0.819 & 0.082 & 0.929 \\
Unseen simulated engines  & per recording & 0.819 & 0.108 & \\
\midrule
\multicolumn{5}{@{}l}{\emph{Test}} \\
Test engine               & per microphone & 0.640 & 0.118 & 0.914 \\
                          & session average & 0.781 & 0.099 & 0.954 \\
Unseen simulated engines  & per recording & 0.785 & 0.114 & 0.938 \\
\bottomrule
\end{tabular}
\end{table}

On the held-out sessions, an estimator trained at the labelled speed and used at the decoded one explains only 0.48 of the variance, so the estimator has to learn under the errors of the speed it receives.

\section{Discussion and Limitations}
\label{sec:discussion}

Reading the load from order levels ties the estimator to the engines and recording chains of its training data. On the unseen-type engine, a four-cylinder recorded with another setup, it follows load ramps at constant speed within a recording only weakly (0.32 of the variance) and misplaces the absolute reading between recordings (none of the pooled variance), whereas a linear fit of the level relative to each recording, made on the training engines, reaches 0.53; one common level offset does not correct the reading. The training data hold a single four-cylinder engine, of another type, and a single recording chain, so engine type and chain cannot be told apart as the cause.

The real material was recorded mostly on chassis dynamometers and holds no road, wind or traffic noise. The reference torque is the engine control unit's own estimate, not a measured shaft torque, and all load errors are stated against it. The conversion to torque in \Nm{} (Eq.~\eqref{eq:torque}) needs the engine's drag floor and full-load ceiling, and its error is not measured here.

\section{Conclusion}
\label{sec:conclusion}

Engine speed and an engine-normalised load can be read from a microphone signal without tachometer or bus access. On a logarithmic speed grid, a relative speed change shifts all orders by one common number of candidates, and tiling kernels collect a harmonic wherever it falls between them. With a decoder that charges changes of the relative rate of speed change, the flat order sum alone places more frames near the true speed than general pitch trackers given the best harmonic order, without any learning. For the load, the order levels at the decoded speed explain 0.86 to 0.95 of the torque variance, whereas the level explains at most 0.52 and the drivetrain equation gives only the sign of the net torque. Recordings without telemetry can thereby be annotated for Active Sound Design and conditioned engine synthesis; the load, so far, only on recording chains like those of the training data.

\bibliographystyle{IEEEtran}
\bibliography{refs}

\end{document}